\documentclass[reprint, superscriptaddress, showpacs,amsmath, amssymb, aps]{revtex4-2}
\usepackage{graphicx}
\usepackage{dcolumn}
\usepackage{bm}
\usepackage{hyperref}
\usepackage{float}
\usepackage{mathrsfs}
\usepackage{amsmath}
\usepackage{slashed}
\usepackage{subcaption}
\usepackage{caption}
\usepackage{comment}
\usepackage{multirow}
\hypersetup{colorlinks=true, citecolor=blue, urlcolor=blue, linkcolor=blue}

\begin{document}
	\title{Investigation of \(\Lambda_c\) States and \((\bar{D}N)\) Molecules Production at EicC and EIC}
	
	\author{Kai-Sa Qiao}\email{qiaokaisa@itp.ac.cn}
	\affiliation{CAS Key Laboratory of Theoretical Physics, Institute of Theoretical Physics, \\
		Chinese Academy of Sciences, Beijing 100190, China}
	\affiliation{School of Physics, University of Chinese Academy of Sciences (UCAS), Beijing 100049, China}
	
	\author{Bing-Song Zou} \email{zoubs@mail.tsinghua.edu.cn}
	\affiliation{Department of Physics, Tsinghua University, Beijing 100084, China}
    \affiliation{CAS Key Laboratory of Theoretical Physics, Institute of Theoretical Physics, \\
		Chinese Academy of Sciences, Beijing 100190, China}
    \affiliation{School of Physics, University of Chinese Academy of Sciences (UCAS), Beijing 100049, China}
   \affiliation{Southern Center for Nuclear-Science Theory (SCNT), \\
   Institute of Modern Physics, Chinese Academy of Sciences, Huizhou 516000, China}

	\date{\today}
	\begin{abstract}
		We explore various \(\Lambda_c\) states, including \(\Lambda_c\), \(\Lambda_c(2595)\), \(\Lambda_c(2940)\), and the predicted \((\bar{D}N)\) hadronic molecular states, in photoproduction and electroproduction to estimate their yields at EicC and EIC. Assuming \(\Lambda_c(2940)\) as either a hadronic molecular state or a three-quark state, our analysis demonstrates that its production rates are of the same order of magnitude, posing challenges in identifying its underlying structure. After considering the integral luminosity, the yields of \(\Lambda_c\) excited states reach \(10^6\) to \(10^7\) at EicC and EIC. The \((\bar{D}N)\) molecular states with both isospin $I = 0$ and $I = 1$ are also studied, with yields reaching $10^5$, making them likely to be detectable at these facilities.
	\end{abstract}
	\pacs{}
	\maketitle
	
	\section{INTRODUCTION}
	
	Searching for hadronic molecule states in charm baryons is an interesting topic in hadron physics. Since the BaBar Collaboration observed $\Lambda_c(2940)$ in the $D^0p$ invariant mass distribution~\cite{babarcollaboration2007}, as a likely candidate for a hadronic molecular state, many groups have studied its nature with different pictures. ~\citet{he2007} proposed that $\Lambda_c(2940)$ is a $D^{*0}p$ molecular state with the quantum number $J^P = 1/2^-$. Using SU(4) effective Lagrangians, ~\citet{dong2010b} suggested $1/2^+$ for  $\Lambda_c(2940)$ and excluded $1/2^-$ since the calculated partial widths are larger than the experimental results. In quark model calculations~\cite{capstick1986a,ebert2011a,chen2017a}, the states closest in mass to \(\Lambda_c(2940)\) are $\Lambda_c(\frac{1}{2}^-,2P)$ and $\Lambda_c(\frac{3}{2}^-,2P)$. Their masses are approximately 40 MeV and 60 MeV higher than that of \(\Lambda_c(2940)\), respectively. In 2020, ~\citet{luo2020} considered the coupled-channel effect by introducing the \(D^*N\) contribution, and found that the mass of \(\Lambda_c(\frac{3}{2}^-, 2P)\) could be consistent with \(\Lambda_c(2940)\). However, in this scenario, the mass of \(\Lambda_c(\frac{1}{2}^-,2P)\) is higher than that of \(\Lambda_c(\frac{3}{2}^-, 2P)\), which has never been observed before. While in Ref.~\cite{Xie2015}, role of the $\Lambda_c(2940)$ in $\pi^- p \to D^- D^0 p$ reaction was theoretical studied within an effective Lagrangian approach.
	
	The study of charm baryons in photoproduction and electronproduction has a long history~\cite{abe1984,abe1986,amendotia,adamovich1987,alvarez1990,na14/2collaboration1993,frabetti1993a,frabetti1994, zeuscollaboration2005, zeuscollaboration2013}. In 1987, NA 1 experiment~\cite{amendotia} at CERN observed 9 $\Lambda_c$ decays. In 1990, the NA14/2 experiment~\cite{alvarez1990} observed \(29 \pm 8 \ \Lambda_c(\bar{\Lambda}_c)\) decays using a mean photon energy of 100 GeV. The E687 experiment~\cite{frabetti1993a} at Fermilab reconstructed 1340 \(\Lambda_c\) decays and measured its lifetime in 1993. They later reported 39.7 \(\pm\) 8.7 excited states \(\Lambda_c(2625)\) in 1994~\cite{frabetti1994}. For electron-production, the ZEUS collaboration at HERA reconstructed \(1440 \pm 220\) \(\Lambda_c\) in 2005~\cite{zeuscollaboration2005} and \(7682 \pm 964\) \(\Lambda_c\) in 2013~\cite{zeuscollaboration2013}, respectively. Various groups have investigated theoretically  the photoproduction of the \(\Lambda_c\) ground state~\cite{rekalo2001,liu2003,tomasi-gustafsson2004,tomasi-gustafsson2005,huang2016a} and \(\Lambda_c(2940)\) as a \(\frac{1}{2}^{\pm }\) molecular state~\cite{wang2015}.
	
	The Electron-Ion Accelerator in China (EicC)~\cite{anderle2021} and the Electron-Ion Collider (EIC)~\cite{khalek2022} in the United States are proposed as complementary to ongoing scientific programs at Jefferson Laboratory. They provide high luminosity at different energy ranges, see Table~\ref{tab:EicCandEIC}. 
	\begin{table*}[htbp]
		\renewcommand{\arraystretch}{1.5}
		\begin{ruledtabular}
			\begin{tabular}{cccc}
				Facility  &  Center-of-Mass Energy(GeV) & Luminosity($\mathrm{cm}^{-2}\cdot\mathrm{s}^{-1}$)& Integrated Luminosity(fb$^{-1}$)\\ \hline
				EicC  &  15-20 & 2$\times 10^{33}$& 50\\ 
				EIC  & 20-140 &   $10^{33-34}$ & 10-100 \\ 
			\end{tabular}
		\end{ruledtabular}
		\caption{Energy, luminosity, and integrated luminosity for EicC and EIC. Integrated luminosity for EicC corresponds to operating time accounting for 80\% of the entire year. Integrated luminosity for EIC corresponds to 30 weeks of operations.}
		\label{tab:EicCandEIC}
	\end{table*}
	Compared to previous photoproduction and electroproduction experiments, their higher luminosity can help us obtain more particle events. Therefore, we can estimate the yields predicted by different models to determine which model the particle belongs to.
	
	In the present work, the process \(\gamma p \rightarrow \Lambda_c^{(*) } \bar{D}^0\) is studied, where \(\Lambda_c^{(*)}\) includes the ground state, \(\Lambda_c(2595)\), and \(\Lambda_c(2940)\). We consider \(\Lambda_c(2940)\) both as a molecular state with $J^P = \frac{1}{2}^{-}$ and as the traditional quark model \(\Lambda_c(\frac{1}{2}^-, 2P)\), and compare the results with those obtained for the ground state and \(\Lambda_c(2595)\) in the quark model to highlight the differences in cross sections between the models. Effective Lagrangians are used to describe these processes, with the hadronic molecule model and the \(^3P_0\) model employed to determine the coupling constants at the vertices. The equivalent photon approximation is then applied to estimate the electronproduction cross section in the EicC and EIC energy regions. 

	Additionally, we also estimate the production rates of the predicted (\(\bar{D}N\)) molecular states~\cite{Peng:2022, yamaguchi2022, Yan:2024} at EicC and EIC to study the feasibility of discovering these molecular states.
	
	This paper is organized as follows. Section 2 outlines the formalisms and foundational elements of the calculations. Section 3 presents the results, covering both photoproduction and electronproduction. The final section offers discussions and a brief summary.
	\section{FORMALISM}
	
	\subsection{The equivalent photon approximation}
	Ultra-relativistic electroproduction can be calculated using the Weizsäcker-Williams' method~\cite{budnev1975}. The cross section for inelastic electron scattering off proton is expressed in terms of cross section $\sigma_\gamma(\omega)$ for the absorption of real photons with frequency $\omega$:
	\begin{equation}
		d\sigma_{ep} = \sigma_\gamma(\omega)dn(\omega, q^2)
	\end{equation}
	The equivalent photon number or spectrum, $dn$, depends on $\omega$ and $q^2$, and its form is defined by the 
	$e\rightarrow e'\gamma^*$ vertex. In the target rest frame, $\omega$ coincides with the photon energy. In numerous cases that $\omega \gtrsim \Lambda_\gamma$, the expressions for the spectrum can be simplified
	\begin{eqnarray}
			\begin{aligned}
			dn (\omega,q^2)&= \frac{\alpha}{\pi}\frac{d\omega}{\omega}\frac{d(-q^2)}{|q^2|}\\
			& \times[1-\frac{\omega}{E}+\frac{\omega^2}{2E^2}-(1-\frac{\omega}{E})|\frac{q^2_{min}}{q^2}|]
		\end{aligned}
	\end{eqnarray}

	After integrating over the region $q_{min}^2\leqslant -q^2 \leqslant q_{max}^2$, the cross section can be written in the form
	\begin{gather}
		d\sigma= \sigma_{\gamma}(\omega)dn(\omega) \\
		dn(\omega) = \int_{q_{min}^2}^{q_{max}^2}dn(\omega,q^2) = N(\omega)\omega d\omega \\
		N(\omega) = \frac{\alpha}{\pi}[(1-\frac{\omega}{E}+\frac{\omega^2}{2E^2})\ln \frac{\Lambda_\gamma^2E(E-\omega)}{m_e^2\omega^2}-(1-\frac{\omega}{E})]
	\end{gather}
	where $\alpha$ is the fine-structure constant, $\Lambda_\gamma$ is a dynamic cutoff, and $\Lambda_\gamma \sim m_\rho$  for the electroproduction off proton. $E$ coincides with electron energy in the target rest frame. \(\omega_{max} = E - 100m_e\) ensures that \(E - \omega \gg m_e\), satisfying the ultra-relativistic condition.

	\subsection{Feynman diagrams and interaction Lagrangian densities}
	The contribution of the t-channel predominates in the photoproduction, hence we only consider t-channel particle exchange here.
	\begin{figure}[htpb]
		\includegraphics{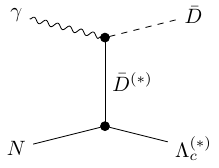}
		\caption{\label{fig:tchannel} Feynman diagram of the t-channel in $\gamma N\rightarrow \bar{D}\Lambda_c$.}
	\end{figure}
	
	We use the following effective Lagrangian to study the $\gamma + p \rightarrow \bar{D}^0+ \Lambda^{(*)+}_c$ reaction.

	\begin{align}
		&\mathcal{L}_{ND\Lambda_c(1/2^+)}= ig_{ND\Lambda_c}\bar{\Lambda}_c\gamma_5ND+H.c., \\
		&\mathcal{L}_{ND^*\Lambda_c(1/2^+)} =g_{ND^*\Lambda_c}\bar{\Lambda}_c\gamma_\mu ND^{*\mu}+H.c., \\
		&\mathcal{L}_{ND\Lambda^*_c(1/2^-)}= ig^{1/2^-}_{ND\Lambda^*_c}\bar{\Lambda}^*_cND+H.c., \\
		&\mathcal{L}_{ND^*\Lambda^*_c(1/2^-)} =g^{1/2^-}_{ND^*\Lambda^*_c}\bar{\Lambda}^*_c\gamma_5\gamma_\mu ND^{*\mu}+H.c., \\
		&\mathcal{L}_{\gamma NN}=-e\bar{N}(Q_N\slashed{A}+\frac{\kappa_N}{4m_N}\sigma^{\mu\nu}F_{\mu\nu})N,\\
		&\mathcal{L}_{\gamma DD} = ieA_\mu(D^+\partial^\mu D^--\partial^\mu D^+D^-), \\
		&\mathcal{L}_{\gamma DD^*} = g_{\gamma DD^*}\epsilon_{\mu\nu\alpha\beta}(\partial^\mu A^\nu)(\partial^\alpha D^{*\beta}) D + H.c.,
	\end{align}
	
	$N, \Lambda_c, \Lambda^*_c,D, D^*$ are nucleon field, $\Lambda_c$ field, $\Lambda_c(2940)$ or $\Lambda_c(2595)$ field, $D$ field and $D^*$ field, respectively. $Q_N$ is nucleon electric charge operator in the unite of $e$. $Q_N = \mathrm{diag}(1,0)$. The anomalous magnetic moment $\kappa_p = 2.792$. The definitions of $\sigma^{\mu\nu}$ and $F^{\mu\nu}$ are:
	\begin{align}
			\sigma^{\mu\nu} &= \frac{i}{2}[\gamma^\mu,\gamma^\nu]\\
			F^{\mu\nu} &= \partial^\mu A^\nu - \partial^\nu A^\mu
	\end{align}

	The coupling constants $g_{ND\Lambda_c} =-13.72$ and $g_{ND^*\Lambda_c} = -5.20$ are obtained using $SU(4)$ flavor symmetry~\cite{dong2010b}. The excited states coupling constants $g_{ND\Lambda_c^*} = -0.54$ and $g_{ND^*\Lambda_c^*} = 6.64$ are used in Refs.~\cite{dong2014, wang2015}. 
	
	The coupling constant $g_{\gamma DD^*}$ can be determined from the reaction $D^* \rightarrow D + \gamma$.
	\begin{equation}
		\Gamma_{D^*\rightarrow D+\gamma} = \frac{g_{\gamma DD^*}^2(m_{D^*}^2-m_D^2)^3}{96\pi m_{D^*}^3}
	\end{equation}
	The value $\Gamma_{D^{*0}\rightarrow D^0 \gamma} = 23\,\mathrm{KeV}$ is derived from the partial decay width $\Gamma(D^{*+}\rightarrow D^0 \pi^+)$ and the decay ratio of  $D^{*0}$~\cite{dong2008,particledatagroup2022}. By further utilizing $m_{D^{*0}} = 2.007\,\mathrm{GeV}$, $m_{D^0} = 1.865\,\mathrm{GeV}$, we can determined $g_{\gamma D^0D^{*0}} = 0.58\,\mathrm{GeV}^{-1}$. \(g_{\gamma D^+D^{*+}} = 0.117\,\mathrm{GeV}^{-1}\) can be directly derived from experimental results.
	
	In the calculations, we need to introduce form factors to suppress the contributions of inner particles when they are strongly off-shell~\cite{lin2017,gao2010}. 
	\begin{gather}
		f_1(q^2) = \frac{\Lambda_1^4}{\Lambda_1^4 + (q^2-m_{ex}^2)^2} \\
		f_2(q^2) = (\frac{\Lambda_2^2 - m_{ex}^2}{\Lambda_2^2 - q^2})^2
	\end{gather}
	$f_1$ and $f_2$ are multipole and monopole form factors, respectively, applied in meson exchange within triangle loops and $t$-channels. Here, \( q \) denotes the four-momentum and \( m \) represents the mass of the exchanged particle.
	
	\subsection{Coupling constants}
	
	\subsubsection{Hadronic molecular structure}
	In this work, we consider both \(\Lambda_c(2940)\) and \(\bar{D}N\) state as $S$-wave molecular states with \(J^P = 1/2^-\) quantum numbers. The effective Lagrangian describing the molecular state and its components, using \(\Lambda_c(2940)\) as an example, is, 
	\begin{equation}
		\begin{aligned}
			\mathcal{L}_{\Lambda_c^*} (x)= &g_{\Lambda_c^*}\bar{\Lambda}_c^*(x)\gamma_5\gamma_\mu\int d^4y \varPhi(y^2)N(x+w_{D^*N}y) \\
			& \times D^{*\mu}(x-w_{ND^*}y)+H.c.
		\end{aligned}
	\end{equation}
	The definition of the kinematical parameters $w_{ij}$ is, 
	\begin{eqnarray}
		w_{ij}  = \frac{m_i}{m_i+m_j}
	\end{eqnarray}
	The correlation function $\varPhi(y^2)$ is used to constrain the components in the vertex~\cite{faessler2007,faessler2007a,faessler2008}. The Fourier transform of the correlation function reads
	\begin{eqnarray}
		\varPhi(y^2)= \int \frac{d^4p}{(2\pi)^4}e^{-ipy}\widetilde{\varPhi}(-p^2)
	\end{eqnarray}
	
	We use the Gaussian form in momentum space to suppress the ultraviolet region.
	\begin{eqnarray}
		\tilde{\varPhi}(p_E^2) \doteq \mathrm{exp}(-p_E^2/\Lambda^2)
	\end{eqnarray}
	where $p_E$ is the Euclidean Jacobi momentum, and $\Lambda$ is a size parameter for the molecular structure.

	The coupling constant $g_{(\bar{D}N)-\bar{D}N}, g_{\Lambda_c^*D^*N}$ with $J^P=1/2^-$ could be determined by the compositeness condition~\cite{Guo:2017jvc},
	\begin{eqnarray}
		Z = 1 - \Sigma'(m^2)=0
	\end{eqnarray} 
	where \(\Sigma'(m^2)\) is the derivative of the mass operator for molecular states.
	
	Considering that $\Lambda_c(2940)$ is described as $\frac{1}{\sqrt{2}}(|D^{0}p\rangle + |D^{+}n\rangle)$, the coupling constant at $\Lambda = 1$ GeV is $g_{\Lambda_c(2940)D^*N} = 1.63$. This  coupling constant changes slowly with variations in $\Lambda$, as shown in Ref.~\cite{Yue:2024paz}.
	
\begin{figure}[htpb]
		\includegraphics[width=0.36\textwidth]{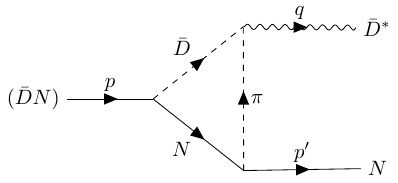}
		\caption{\label{fig:dia_Triangle_loop} Diagram of \((\bar{D}N) - \bar{D}^*N\) vertex with meson exchange}
	\end{figure}

	Coupling constants between molecular states and other channels are determined via form factors in triangle loop diagrams. In the calculation of the photoproduction process involving the \(\bar{D}N\) molecular state, we also account for \(D^*\) exchange, necessitating consideration of the coupling constant at the \((\bar{D}N) - \bar{D}^*N\) vertex (Fig.~\ref{fig:dia_Triangle_loop}). We consider the exchange of $\pi$ mesons to estimate the photoproduction cross section.
	
	The Lagrangians used here include the VPP vertex in SU(4) and the \(\pi NN\) vertex.
	\begin{align}
		\mathcal{L}_{\pi NN} &=-\frac{g_{A}}{2f_{\pi}}\bar{N}\gamma^\mu\gamma_5\partial_\mu\vec{\pi}\cdot\vec{\tau}N \\
		\mathcal{L}_{VPP} &= ig_{VPP}\langle[\partial_\mu P,P]V^\mu\rangle
	\end{align}
	
	In the above Lagrangians, the coupling constants are set as follows: \(g_{VPP} = 4.67\), \(g_A = 1.267\), and \(f_\pi = 92.4\) MeV. \(\tau_a\) represents the Pauli matrices, while \(P\) and \(V^\mu\) denote the 16-plet of pseudoscalar and vector fields, respectively. \(N\) represents the nucleon field.
	
	The Lagrangian of \((\bar{D}N)-\bar{D}^*N\) consists of two parts and we use \(B'\) and \(B\) to represent \((\bar{D}N)\) and \(N\).
	\begin{equation}
		\begin{aligned}
			\mathcal{L}_{B'BV} =& \bar{B'}_1(g_{B'BV}\gamma_5\gamma_\mu+\frac{f_{B'BV}}{m_1-m_2}\gamma_5\sigma^{\mu\nu}\partial_\nu)V_\mu B_2 \\
			&+ H.c.
		\end{aligned}
	\end{equation} 
	
	We represent the vertex in Fig.~\ref{fig:dia_Triangle_loop} as \(\Gamma^\mu(q)\). Thus, we have:
	\begin{gather}
		\gamma_5\Gamma^\mu(q^2) = \gamma_5[F_1(q^2)\gamma^\mu + F_2(q^2)p^\mu+ F_3(q^2)p'^\mu] \\
		g_{B'BV} = F_1-\frac{1}{2}(F_2+F_3)(m_1-m_2) \\
		f_{B'BV} =-\frac{1}{2}(F_2+F_3)(m_1-m_2) 
	\end{gather}
	
	\subsubsection{$^3P_0$ model}
	The \( ^3P_0 \) model, also referred to as the quark pair creation model~\cite{roberts1992,blundell1996,chen2007}, is utilized here to study the coupling constant at the \( A \rightarrow BC \) vertex, treating all involved particles as composite states. Define the S matrix
	\begin{align}
		S &\equiv I - 2\pi i\delta(E_f-E_i)T\label{eq:S}\\
		\langle f|T|i\rangle &\equiv \delta^{(3)}(\mathbf{P}_f - \mathbf{P}_i)M^{M_{J_A}M_{J_B}M_{J_C}}\label{eq:T}
	\end{align}
	$T$ is the transition operator, defined as
	\begin{equation}
	\begin{aligned}
		T = &-3\gamma \sum_m \langle 1m,1-m|00\rangle\int d^3 p_4d^3p_5\delta^{(3)}(\vec{p}_4+\vec{p}_5)\\
		&\times \mathcal{Y}_1^m(\frac{\vec{p}_4-\vec{p}_5}{2})\chi_{1-m}^{45}\phi_0^{45}\omega_0^{45}b_4^\dag(\vec{p}_4)d_5^\dag(\vec{p}_5)
	\end{aligned}
	\end{equation}
	where $\gamma$ is an undetermined parameter in model, descrping the quark pair production strength from vacuum. $\mathcal{Y}_l^m(p,\theta.\phi) = p^lY_l^m(\theta,\phi)$ is a solid harmonic polynomial denoting the spatial wave funtion in momentum space. $\chi_{0}^{45}$, $\phi_0^{45} = (u\bar{u}+d\bar{d}+s\bar{s})/\sqrt{3}$ , and $\omega_0^{45} = \delta_{ij}$ are spin, flavor and color wave functions repectively. 
	
	\begin{figure}[htbp]
		\centering
		\begin{minipage}[b]{0.15\textwidth}
			\centering
			\includegraphics[width=\textwidth]{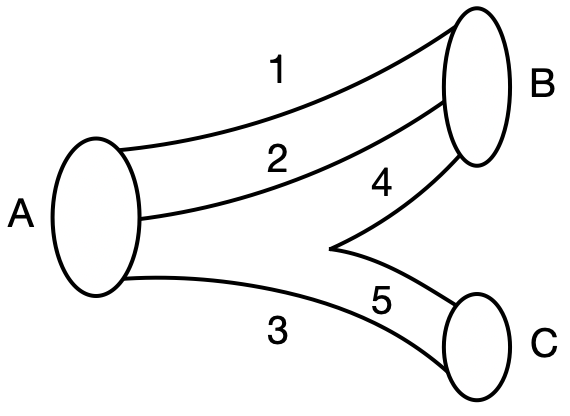}
			\subcaption{}
			\label{fig:3p01}
		\end{minipage}
		\hfill
		\begin{minipage}[b]{0.15\textwidth}
			\centering
			\includegraphics[width=\textwidth]{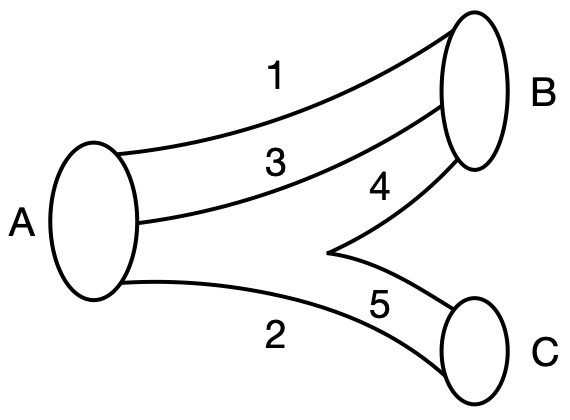}
			\subcaption{}
			\label{fig:3p02}
		\end{minipage}
		\hfill
		\begin{minipage}[b]{0.15\textwidth}
			\centering
			\includegraphics[width=\textwidth]{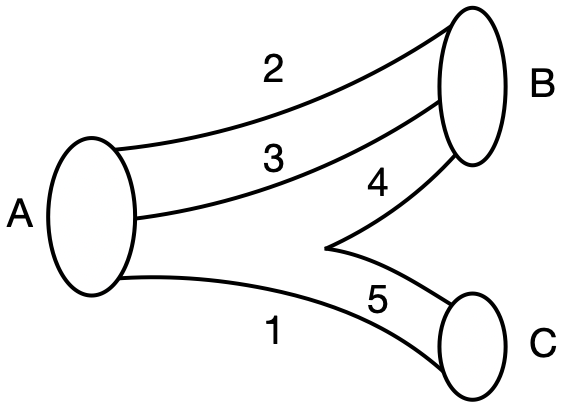}
			\subcaption{}
			\label{fig:3p03}
		\end{minipage}
		\caption{The Vertex $A\rightarrow B+C$ in the $^3P_0$ model}
		\label{fig:3p0}
	\end{figure}
	
	We use mock state~\cite{hayne1982} to define the baryon and meson.
	\begin{equation}
		\begin{aligned}
			&|A(n_A^{2S_A+1}L_AJ_AM_{J_A})(\vec{P}_A)\rangle \\
			&\equiv  \sqrt{2E_A}\sum_{M_{L_A},M_{S_A}}\langle L_AM_{L_A}S_AM_{S_A}|J_AM_{J_A}\rangle \\
			& \times\int d^3p_1d^3p_2d^3p_3 \delta^{(3)}(\vec{p}_1+\vec{p}_2+\vec{p}_3-\vec{P}_A) \phi^{123}_A\omega_A^{123}  \\
			&\times\psi_{n_AL_AM_{L_A}}(\vec{p}_1,\vec{p}_2,\vec{p}_3)\chi^{123}_{S_AM_{S_A}}|q_1(\vec{p}_1)q_2(\vec{p}_2)q_3(\vec{p}_3)\rangle 
		\end{aligned}
	\end{equation}

	The normalization condition of mock state is
	\begin{equation}
		\langle A(\vec{p}_A)|A(\vec{p}_A') \rangle= 2E_A\delta^3(\vec{p}_A-\vec{p}_A')\label{eq:normalization}
	\end{equation}
	
	In our work, we use the $^3P_0$ model to calculate the coupling constants for vertices such as  $\Lambda_c^*\rightarrow ND^{(*)}$. Thus only Fig.~\ref{fig:3p02} meet the condition when the third quark in A is a heavy quark. In fact, these two diagrams are the same, which means we only need to calculate one amplitude. The amplitude of Fig.~\ref{fig:3p02} 
	\begin{equation}
		\begin{aligned}
			&\mathcal{M}^{M_{J_A}M_{J_B}M_{J_C}}(A\rightarrow BC) \\
			& = \gamma\sqrt{8E_AE_BE_C} \sum_{\substack{M_{L_A},M_{S_A},\\M_{L_B},M_{S_B},\\M_{L_C},M_{S_C},m}}\langle L_A M_{L_A}S_AM_{S_A}|J_A M_{J_A}\rangle \\
			&\times \langle L_B M_{L_B}S_BM_{S_B}|J_B M_{J_B}\rangle\langle L_C M_{L_C}S_CM_{S_C}|J_C M_{J_C}\rangle \\
			&\times\langle 1m;1-m|00\rangle \langle\chi_{S_BM_{S_B}}^{134}\chi_{S_CM_{S_C}}^{25}|\chi_{S_AM_{S_A}}^{123}\chi_{1-m}^{45}\rangle\\
			&\times \langle \phi_B^{134}\phi_C^{25}|\phi_A^{123}\phi_0^{45}\rangle I_{M_{L_B}M_{L_C}}^{M_{L_A},m}(\vec{p})
		\end{aligned}
	\end{equation}
	
\begin{figure}[htpb]
		\includegraphics[width=0.13\textwidth]{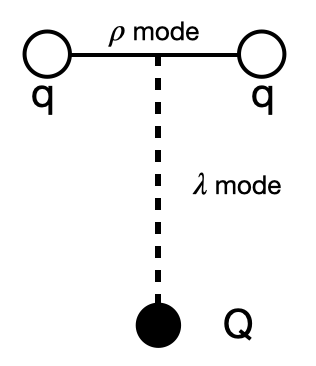}
		\caption{\label{fig:exctitationmode} The excitation mode in heavy baryons: q and Q denote the light and heavy quarks, respectively.}
	\end{figure}

	In the present work, we use simple harmonic oscillator (SHO)~\cite{capstick1986} spatial wave functions to describe both baryons and mesons. This approach provides a good approximation for studying decay properties. The spatial wave function of the baryon consists of two parts, as shown in Fig.~\ref{fig:exctitationmode}, and each part is in the form of an SHO.
	\begin{equation}
		\psi_A(\vec{p}) = N\psi_{n_{\rho}l_{\rho}m_{l\rho}}(\vec{p}_\rho)\psi_{n_\lambda l_\lambda m_{l\lambda}}(\vec{p}_\lambda) 
		\label{eq:hadronSpatial}
	\end{equation}
	
	In Refs.~\cite{blundell1996,yu2023,chen2007}, the parameter $\gamma = 13.4$ was determined by fitting 28 meson decay channels~\cite{blundell1996}. Refs.~\cite{lu2018, lu2020} used $\gamma = 9.83$, derived from fitting the $\Sigma_c(2520)^{++}\rightarrow \Lambda_c + \pi^+$ process. We adopt the latter value, as it is more pertinent to the process under consideration. We assume the following relationship~\cite{lu2018} for the harmonic oscillator parameters in the baryon spatial wave function
	\begin{equation}
		\alpha_\lambda = (\frac{3m_Q}{2m_q+m_Q})^{1/4}\alpha_\rho	
	\end{equation}
	where $\alpha_\rho = 0.4$ GeV. $m_Q$ and $m_q$ are the masses of the heavy and light quarks, respectively, with $m_u = m_d = 220$ MeV, $m_s = 419$ MeV, and $m_c = 1628$ MeV. The harmonic oscillator parameters in mesons are $R = 2.5\, \mathrm{GeV}^{-1}$ for light mesons, $R = 1.67\, \mathrm{GeV}^{-1}$ for $D$ mesons, and $R = 1.94\, \mathrm{GeV}^{-1}$ for $D^*$ mesons~\cite{godfrey2016,xiao2018a,lu2018}.
	
	Since not all particles are on-shell when calculating the coupling constants at the vertex, and the internal line particles are often mesons, we set $p_A^2 = m_A^2$, $p_B^2 = m_B^2$, $p_C^2 = 0$, where $p_C$ is the four-momentum. We derive the effective coupling constants by comparing the $^3P_0$ model amplitude with the one obtained from the effective Lagrangian without the coupling constants.
	\begin{equation}
		g_{ABC} =\sqrt{ \frac{\sum_{\substack{spins}}|\mathcal{M}_{^3P_0}(m_A^2,m_B^2,0)|^2}{\sum_{\substack{spins}}|\mathcal{M}'_{\mathcal{L}}(m_A^2,m_B^2,0)|^2}(2\pi)^3}
	\end{equation}
	The factor  $(2\pi)^3$ stems from the normalization difference between the two corresponding amplitudes in Eqs.(\ref{eq:S},\ref{eq:T},\ref{eq:normalization}).
	
	\section{NUMERICAL RESULTS}
	\subsection{$\Lambda_c$ states}
	
	In this section, we give the numerical results of photoproduction cross section in process $\gamma p \rightarrow \Lambda_c^{(*)} \bar{D}^0$. Charm baryons include $\Lambda_c$, $\Lambda_c(2595)$, $\Lambda_c(2940)$ in the \(^{3}P_0\) model and the hadronic molecule model with $J^P = 1/2^-$. Considering that the center-of-mass energy  is much higher than the nucleon mass, and without taking into account the contributions from heavy resonances, the \( t \)-channel contribution should be dominant. Therefore, we use only the t-channel for the estimation.
	
	The mass of \(\Lambda_c(2940)\) in the quark model is not well-determined. However, the mass of \(\Lambda_c(\frac{1}{2}^-, 2P)\) is close to that of \(\Lambda_c(2940)\), so we use this assignment along with the experimental mass~\cite{babarcollaboration2007} in our $^3P_0$ model calculations.
	
	 In our work, we first calculate the coupling constants for the vertices \(\Lambda_c(2940)-D^*p\) and \(\Lambda_c(2595)-D^*p\) using the \(^3P_0\) model. Since these vertices cannot observed in decay processes, we consider the mesons in these vertices to be off-shell. In this approach, the coupling constant is expressed as a function of momentum, denoted as \(g(p_m^2)\).
	 
	When \(p_m^2 = 0\), the corresponding coupling constants are as follows.
	 \begin{gather}
	 	g_{\Lambda_c(2940)D^*p}^{QPC} = 0.75 \\
	 	g_{\Lambda_c(2595)D^*p}^{QPC} = 1.21
	 \end{gather}
	
	There is only one free parameter, \(\gamma\), in this process. It is determined by fitting the \(\Sigma_c(2520)^{++} \rightarrow \Lambda_c + \pi^+\) decay using the \(^{3}P_0\) model. For simplicity, the Jacobi momentum in the hadron spatial wave function as in Eq.~(\ref{eq:hadronSpatial}) is assumed to be independent of quark mass. 
	
\begin{figure}[htpb]
		\includegraphics[width=0.45\textwidth]{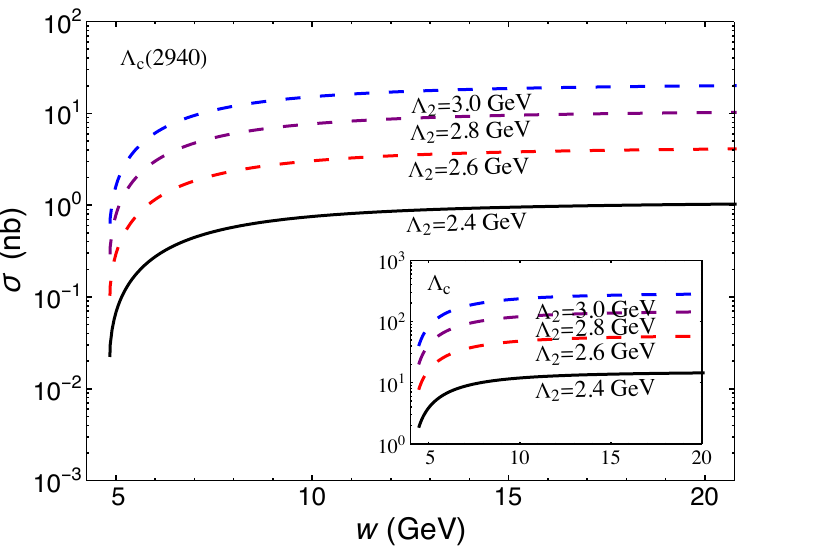}
		\caption{\label{fig:cutoff1} Cross sections of \(\Lambda_c\) and \(\Lambda_c(2940)\) for different cutoff parameters $\Lambda_2$.}
	\end{figure}

	The cross section from the \(t\)-channel meson exchange is sensitive to the cutoff parameter $\Lambda_2$. Empirically \(\Lambda_2\) should be larger than \(m_{ex}\) by $0.4\sim 1.0$ GeV. Here \(m_{ex} = m_{D^{*0}} = 2.007\) GeV, so it is reasonable to take $\Lambda_2$ in the range of $2.4\sim 3.0$ GeV. We use $\Lambda_c$ and $\Lambda_c(2940)$ molecular state as examples, and the results are shown in Fig.~\ref{fig:cutoff1}. As observed, the results increase gradually as the cutoff value rises, but the rate of increase slows down progressively.
	
\begin{figure}[htpb]
			\includegraphics[width=0.45\textwidth]{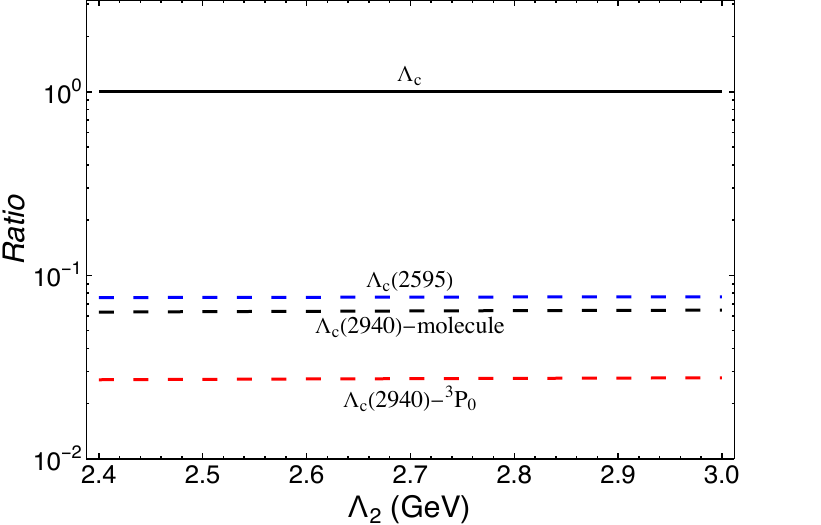}
			\caption{\label{fig:cutoffratio} Ratios of various channels to $\Lambda_c$ for different cutoffs with $w=10$ GeV.}
	\end{figure}

	By comparing the charm production results~\cite{adamovich1987,alvarez1990,na14/2collaboration1993,tomasi-gustafsson2005} and the mass of the exchanged particle in the t-channel, we consider \(\Lambda_2 = 2.5\) GeV to be reasonable. However, the ratios of the excited state channels to the ground state stay stable as the cutoff varies from 2.4 to 3 GeV, as shown in Fig.~\ref{fig:cutoffratio}. 
	
\begin{figure}[htbp]
		\includegraphics[width=0.45\textwidth]{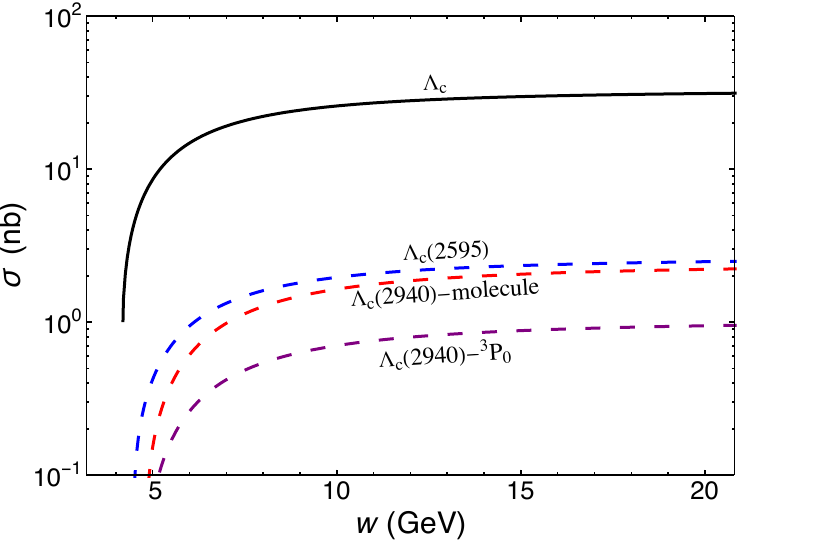}
		\caption{\label{fig:result2} Photoproduction cross sections of \(\Lambda_c\), \(\Lambda_c(2595)\), and \(\Lambda_c(2940)\) in the \(^{3}P_0\) model and the hadronic molecule model within the \(t\)-channel of the process \(\gamma p \rightarrow \bar{D}^0 \Lambda_c^{(*)}\). The cutoff is set to $\Lambda_2=2.5$ GeV.}
	\end{figure}

	The photoproduction results are illustrated in Fig.~\ref{fig:result2}. The cross sections for the three-quark model states decrease as the particle mass increases. In contrast, the hadronic molecule model shows an enhancing effect on \(\Lambda_c(2940)\). However, since the cross sections remain within the same order of magnitude as those predicted by the quark model, distinguishing the structure of \(\Lambda_c(2940)\) through photoproduction alone may not be very decisive.
	
	Electrons in EicC and EIC have high energy, thus we can use the Weizsäcker-Williams' method to calculate electroproduction cross section. The estimation of yields for different states is shown in Table~\ref{table:Yield}. The yields of \(\Lambda_c\) excited states, calculated using integrated luminosity, reach \(10^6\) to \(10^7\) at EicC and EIC. Therefore, even after taking into account of reconstruction efficiency, the yields remain considerable large.

	\begin{table}[htbp]
		\renewcommand{\arraystretch}{1.5}
		\caption{Estimated yields for the states $\Lambda_c$, $\Lambda_c(2595)$, and $\Lambda_c(2940)$ at EicC and EIC.}
		\begin{ruledtabular}
				\begin{tabular}{ccc}
				State & EicC & EIC \\
				\hline
				$\Lambda_c$ & ($6.3\sim9.3$) $\times 10^7$ & $(1.9 \sim 8.0)\times 10^8$ \\ 
				$\Lambda_c(2595)$ & $(4.3\sim 6.6)\times 10^6$ & $(1.3\sim 6.1)\times 10^7$ \\
				$\Lambda_c(2940)$-molecule & $(3.3\sim 5.2) \times 10^6$ & $(1.1 \sim 5.3)\times 10^7$ \\
				$\Lambda_c(2940)$-$^3P_0$ & $(1.4\sim2.2)\times 10^6$ & $(4.5\sim 2.3)\times 10^7$
			\end{tabular}
		\end{ruledtabular}
	\label{table:Yield}
	\end{table}

	\subsection{$(\bar{D}N)$ states}
	
	Photoprocution of $(\bar{D}N)$ processes we included are \(\gamma + p \rightarrow (\bar{D}N) + D^+\). For simplicity, we considered the two states with \(I=0\) and \(I=1, I_3=0\) in the calculation, since both consist of the \((\bar{D}^0n)\) and \((D^-p)\) states. Similar to the case of \(\Lambda_c(2940)\) photoproduction, we only consider the \(t\)-channel here, with \(D\) and \(D^*\) exchange.
	
	\begin{gather}
		|(\bar{D}N),I = 0\rangle = \frac{1}{\sqrt{2}}(|D^-p\rangle - |\bar{D}^0n\rangle) \\
		|(\bar{D}N),I = 1, I_3 = 0\rangle = \frac{1}{\sqrt{2}}(|D^-p\rangle + |\bar{D}^0n\rangle)
	\end{gather}
	
	First, we provide the coupling constants for the vertices \((\bar{D}N)-\bar{D}N\) and \((\bar{D}N)-\bar{D}^*N\). Note that in Ref.~\cite{yamaguchi2022},  two  \((\bar{D}N)\) bound states of \(I=0\) and \(I=1\) are proposed with masses \(m_{(\bar{D}N)}^{I=0} = 2804.8\) MeV and \(m_{(\bar{D}N)}^{I=1} = 2800.2\) MeV, respectively . Here, we take the average of nucleon and \(D\) meson masses. The mass difference due to the isospin symmetry breaking has little effect on the coupling constants, so we present only the averaged results. The cutoff is set to be \(\Lambda = 1\) GeV. 
	\begin{gather}
		g_{(\bar{D}N)}^{I=0} = 1.68 \\
		g_{(\bar{D}N)}^{I=1} =  2.62
	\end{gather}
	The coupling constant of the \(I = 1\) state is higher than that of the \(I = 0\) state due to its deeper binding energy. 
	
	We include \(\pi\) exchange in the triangle loop when calculating the \((\bar{D}N)-\bar{D}^*N\) coupling for estimating its photoproduction rate. The results are shown in Table~\ref{table:DNconstants}. 
	\begin{table}[htbp]
		\renewcommand{\arraystretch}{1.5}
		\caption{Coupling constants in $(\bar{D}N)-\bar{D}^*N$ vertex. The cutoff is set to $\Lambda = 1$ GeV and $\Lambda_1 = 1$ GeV.}
		\begin{ruledtabular}
			\begin{tabular}{cccc}
				States & I= 0 & I= 1 \\
				\hline
				$g_{(\bar{D}N)-\bar{D}^{*}N}$ & 0.40 & -0.21 \\
				$f_{(\bar{D}N)-\bar{D}^{*}N}$& 0.45 &  -0.24
			\end{tabular}
		\end{ruledtabular}
		\label{table:DNconstants}
	\end{table}
    
	The amplitudes of the two components in $(\bar{D}N)$ have different signs in the triangle loops, which makes the couplings in the $I = 0$ state higher than those in the $I = 1$ state.
	
\begin{figure}[htbp]  
		\centering
		\includegraphics[width=0.5\textwidth]{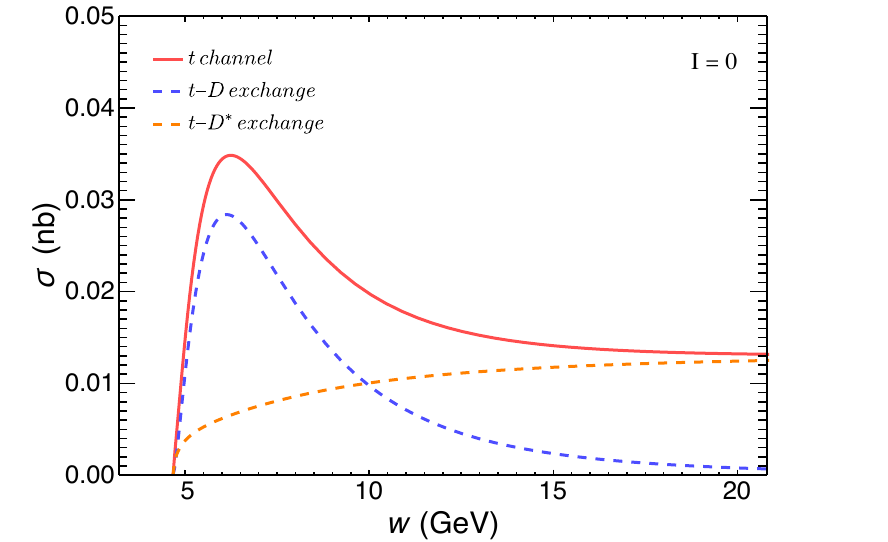}  
		\includegraphics[width=0.5\textwidth]{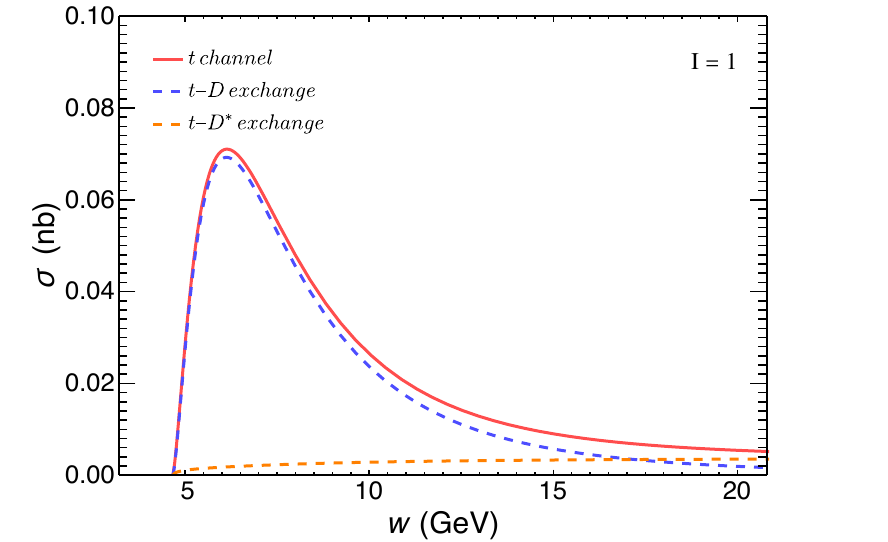}
		\caption{\label{fig:DNresult} Cross sections of \(\gamma + p \rightarrow (\bar{D}N)_I+ D^+\)  for the $\bar{D}N$ molecules with isospin \(I=0\) and \(I=1\). }
	\end{figure}

	Photoproduction cross sections with different channels are shown in Fig.~\ref{fig:DNresult}. The contribution of $D$ meson exchange is similar in both processes, primarily concentrated in the region below 10 GeV. Due to the characteristics of the amplitude structure, \(D^*\) exchange becomes dominant as the energy increases. In the \(I = 1\) process, the suppression of \(g_{(\bar{D}N)-\bar{D}^{*}N}\) causes the onset of this dominance to shift to a higher energy region.

	The estimated yields for the different states are presented in Table~\ref{table:YieldDN}. The yields of the $I = 1$ state are slightly higher than those of the $I = 0$ state, but both remain within the same order of magnitude. However, they are nearly one order of magnitude lower than the yields of the $\Lambda_c$ states. 

	\begin{table}[htbp]
		\renewcommand{\arraystretch}{1.5}
		\caption{\label{table:YieldDN}Estimated yields for the state \((\bar{D}N)\) in different isospin configurations at EicC and EIC.}
		\begin{ruledtabular}
			\begin{tabular}{cccc}
				State & Isospin & EicC & EIC \\
				\hline
				\multirow{2}{*}{$(\bar{D}N)$} & $I = 0$ & $(7.5 \sim 9.8)\times 10^4$ &  $(2.0\sim5.3)\times 10^5$\\ 
				 							& $I = 1, I_3 = 0$ & $(1.3 \sim 1.6)\times 10^5$ &  $(3.2\sim 6.0)\times 10^5$\\
			\end{tabular}
		\end{ruledtabular}
	\end{table}
	
    This difference arises from two sources: the disparity between the coupling constants \(g_{\gamma D^0D^{*0}}\) and \(g_{\gamma D^+D^{*+}}\), and the difference in the dominant contribution channels for the two molecular states. For \(\Lambda_c(2940)\), \(D^*\) exchange plays a significant role, leading to its production being concentrated in the high-energy region. In contrast, for \((\bar{D}N)\), \(D\) exchange dominates, resulting in its production being concentrated in the low-energy region. As a result, the yields difference for \((\bar{D}N)\) between EicC and EIC is not significant, while \(\Lambda_c(2940)\) shows a difference of an order of magnitude.
	 
	\section{SUMMARY}
	In this work, we investigate the \(\Lambda_c\) states, including \(\Lambda_c\), \(\Lambda_c(2595)\), \(\Lambda_c(2940)\), and \((\bar{D}N)\) states, in both photoproduction and electroproduction processes to estimate their yeilds at EicC and EIC. We use effective Lagrangians to describe the amplitudes and apply $^3P_0$ model and hadronic molecule model to calculate the coupling constants. The equivalent photon approximation, the Weizsäcker-Williams method in this work, is used to derive electroproduction from real photon production.
    
	The \(\Lambda_c(2940)\) is studied in both the hadronic molecule model, assigned \(J^P = \frac{1}{2}^-\), and the quark model as the \(\Lambda_c(\frac{1}{2}^-, 2P)\) state. By comparing the four \(\Lambda_c\) states, the hadronic molecule shows an enhancement in its production. However, since it remains within the same order of magnitude, distinguishing the structure of \(\Lambda_c(2940)\) may not be very decisive. Considering the integrated luminosity, the yields of \(\Lambda_c\) excited states are estimated to reach \(10^6\) to \(10^7\) at EicC and EIC.
	
	Production yields of the predicted \((\bar{D}N)\) molecular states with both isospin \(I = 0\) and \(I = 1\) are also estimated. Due to differences in the coupling constants and dominant contribution channels, the yields of the \((\bar{D}N)\)  molecules are approximately one order of magnitude lower than that of \(\Lambda_c(2940)\), but still detectable at EicC and EIC, with yields reaching \(10^5\).

    Our investigations demonstrate that both EicC and EIC will provide great opportunities to study charm baryons of various configurations.

\bigskip


	\begin{acknowledgments}
		We thank Feng-Kun Guo, Jia-Jun Wu, Yong-Hui Lin, and Shu-Ming Wu for their useful discussions and valuable comments. This work is supported by the NSFC and the Deutsche Forschungsgemeinschaft (DFG, German Research Foundation) through the funds provided to the Sino-German Collaborative Research Center TRR110 “Symmetries and the Emergence of Structure in QCD” (NSFC Grant No. 12070131001, DFG Project-ID 196253076 - TRR 110), by the NSFC Grant No.11835015, No.12047503, and by the Chinese Academy of Sciences (CAS) under Grant No.XDB34030000.
	\end{acknowledgments}

	\bibliography{Papers}
	
\end{document}